\documentclass[12pt,prd,tightenlines,nofootinbib,showpacs,showkeys]{revtex4}
\newcommand{\be}{\begin{equation}}
\newcommand{\ee}{\end{equation}}
\newcommand{\bdis}{\begin{displaymath}}
\newcommand{\edis}{\end{displaymath}}
\newcommand{\bga}{\begin{equation}\begin{gathered}}
\newcommand{\ega}{\end{gathered}\end{equation}}
\usepackage{bm}
\usepackage{graphics}
\usepackage{rotating}
\usepackage{epsfig}
\usepackage{amsmath}
\usepackage{amsfonts}

\begin{document}

\title{Radiative nonrecoil nuclear finite size corrections of order
$\alpha(Z\alpha)^5$ to the Lamb shift in light muonic atoms}
\author{\firstname{R.~N.} \surname{Faustov}}
\affiliation{Institute of Informatics in Education, FRC CSC RAS, Vavilov Str. 40, 119333, Moscow, Russia}
\author{\firstname{A.~P.} \surname{Martynenko}}
\author{\firstname{F.~A.} \surname{Martynenko}}
\author{\firstname{V.~V.} \surname{Sorokin}}
\affiliation{Samara University, Moskovskoye Shosse 34, 443086,
Samara, Russia}

\begin{abstract}
On the basis of quasipotential method in quantum electrodynamics
we calculate nuclear finite size radiative corrections of order
$\alpha(Z\alpha)^5$ to the Lamb shift in muonic hydrogen and helium. 
To construct the interaction potential of particles, which gives the necessary 
contributions to the energy spectrum, we use the method of projection operators to states 
with a definite spin. Separate analytic expressions for the contributions 
of the muon self-energy, the muon vertex operator and the amplitude with
spanning photon are obtained. We present also numerical results for these
contributions using modern experimental data on the electromagnetic form 
factors of light nuclei.
\end{abstract}



\maketitle

The investigation of the Lamb shift and hyperfine structure of the muonic hydrogen 
and helium spectrum opened a new page of studies of the energy spectra of simplest atoms.
The experiments that are currently being carried out by the collaboration 
CREMA (Charge Radius Experiments with Muonic Atoms) \cite{crema1,crema2,crema3,crema4}
will make it possible to additionally test the Standard model, obtain more accurate values for 
a number of fundamental parameters, and possibly answer the question of the presence 
of additional exotic interactions between the particles.
The inclusion of other experimental groups in this field of research 
(see, \cite{ma_2017,adamczak_2017,pohl_2017}) will allow, 
as planned, not only to verify the experimental results of the CREMA collaboration, 
but also to lead to a further increase in the accuracy of the experimental 
results for separate intervals of fine and hyperfine structure.
The already obtained results of the CREMA collaboration show that there is a significant 
difference between the values of such a fundamental parameter as the charge radius 
of the nucleus obtained from the study of electronic and muonic atoms.
As has always been the case for a long history of precision studies of the energy spectra 
of simplest atoms in quantum theory, one of the ways to overcome the crisis situation 
is related to a new more in-depth theoretical analysis, recalculation of various theoretical 
contributions that can be amplified in the case of muonic atoms.
In this way, the problem of a more accurate theoretical construction of the particle interaction 
operator in quantum electrodynamics, the calculation of new corrections in the energy spectrum 
of muonic atoms acquires a special urgency \cite{apmlet}.

\begin{figure}[t!]
\centering
\includegraphics{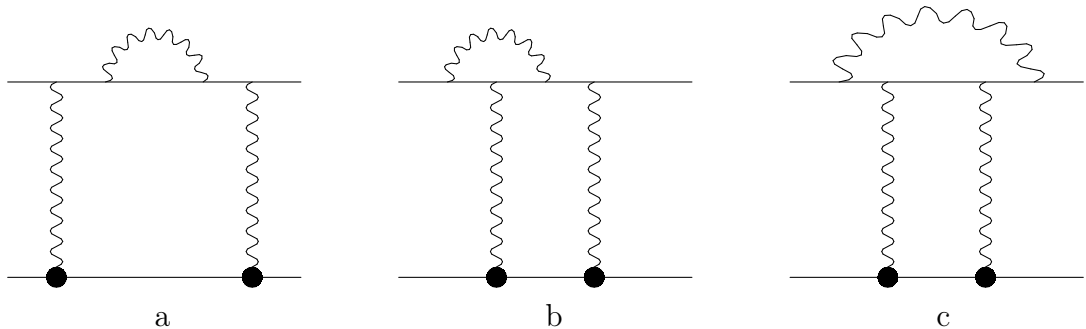}
\caption{Direct two-photon exchange amplitudes with radiative corrections to muon line
giving contributions of order $\alpha(Z\alpha)^5$ to the Lamb shift. Wave line on the diagram represents
the photon. Bold point on the diagram represents the nucleus vertex operator.}
\label{fig1}
\end{figure}

In this work we study radiative nonrecoil corrections of special kind of order $\alpha(Z\alpha)^5$ 
related with the finite size of the nucleus in the Lamb shift of muonic hydrogen and helium. 
A preliminary estimate of the possible magnitude of such a contribution to the Lamb shift can be 
obtained on the basis of general factor $\alpha^6\mu^3/m_1^2\approx 0.012$ meV (for the $\mu p$). 
It shows that the contributions of this order should be studied more closely.
While the theoretical contribution of a certain order is not calculated accurately, the 
theoretical error caused by it is retained, which, when estimated by the main factor, 
can reach a considerable value. For precise determination
of order $\alpha(Z\alpha)^5$ contribution we should account that the distribution of the nucleus
charge is described by the nucleus electric form factor. Radiative nonrecoil corrections of
order $\alpha(Z\alpha)^5$ are divided into three parts: muon self-energy correction, vertex correction
and correction with spanned photon presented in Fig.\ref{fig1}.
This work continues the investigation of the radiative nonrecoil corrections of order $\alpha(Z\alpha)^5$ 
made earlier in our works \cite{apmplb,apm2014} to the case of the Lamb shift.

It is necessary to point out that the contribution of amplitudes shown in Fig.\ref{fig1}
in the case of the point nucleus was calculated many years ago in \cite{kks}. 
Taking into account the finite size of the nucleus, the contribution of these diagrams to the Lamb shift
of hydrogen atom proportional to the square of charge radius $<r_N^2>$  was obtained numerically in \cite{kp1993,eides1997}.
In analytical form the contribution of amplitudes in Fig.~\ref{fig1} 
proportional to $<r_N^2>$ was obtained in \cite{milstein1,milstein2}. In this study
we obtain closed integral expressions for the contributions of individual diagrams, and then calculate them analytically in the approximation, when the electric form factor of the nucleus is replaced by the square of the charge radius, and numerically, taking into account the complete expression of the form factor obtained on the basis of experimental data. We have presented the required corrections in such a clear form that can be used 
in the future to evaluate numerically the contributions of the most diverse simple atoms.

To study the Lamb shift of muonic atom, we use a quasipotential method in quantum electrodynamics 
in which the bound state of a muon and nucleus is described in the leading order 
in the fine-structure constant by the Schr\"odinger equation with the Coulomb potential \cite{apmjetp,apm2005,apm1999}. 
The first part of important corrections in the energy spectrum 
is determined by the Breit Hamiltonian \cite{apmjetp,apm2005,apm1999}. 
Other corrections can be obtained by studying the different interaction amplitudes of particles
(see the detailed review of Eides, Grotch and Shelyuto \cite{egs}).
To evaluate radiative nonrecoil corrections of order $\alpha(Z\alpha)^5$ we neglect relative momenta of particles
in initial and final states and construct separate potentials corresponding to muon self-energy,
vertex and spanning photon diagrams in Fig.\ref{fig1}.

The contribution of two-photon exchange diagrams to the Lamb shift and hyperfine structure 
of order $(Z\alpha)^5$ was investigated earlier by many authors \cite{egs,borie1,borie2,kp1996,tomalak}. 
The lepton line radiative corrections to two-photon exchange amplitudes
were studied in detail in \cite{eides1,eides2} including recoil effects. 
Numerous studies have shown that corrections of this type are conveniently calculated
in the Fried-Yennie gauge for radiative photon \cite{fried} because
it leads to infrared-finite renormalizable integral expressions for muon self-energy operator, 
vertex function and lepton tensor describing the diagram with spanning photon.
It is convenient to begin the study of corrections $\alpha(Z\alpha)^5$ from the most general 
expression for the amplitudes (direct and crossed) shown in Fig.~\ref{fig1}.
In order to demonstrate the calculation technique, let us consider a direct and a crossed amplitude 
with arbitrary muon tensor insertion omitting a number of nonessential factors:
\begin{equation}
\label{eq:2gamma1}
{\cal T}_{direct}=\left[
\bar u(q_1)L_{\mu\nu}u(p_1)\right]
\Bigl[\bar v(p_2)\Bigl(\gamma_\lambda F_1-\frac{1}{2m_2}\sigma_{\lambda\rho}k_\rho F_2\Bigr)
\frac{-\hat p_2-\hat k+m_2}{(p_2+k)^2-m_2^2}\times
\end{equation}
\begin{displaymath}
\Bigl(\gamma_\sigma F_1+\frac{1}{2m_2}\sigma_{\sigma\beta}k_\beta F_2\Bigr)v(q_2)\Bigr]
D_{\mu\sigma}(k)D_{\nu\lambda}(k),
\end{displaymath}
\begin{equation}
\label{eq:2gamma2}
{\cal T}_{crossed}=\left[
\bar u(q_1)L_{\nu\mu} u(p_1)\right]
\Bigl[\bar v(p_2)\Bigl(\gamma_\lambda F_1+\frac{1}{2m_2}\sigma_{\lambda\rho}k_\rho F_2\Bigr)
\frac{-\hat p_2+\hat k+m_2}{(p_2-k)^2-m_2^2}\times
\end{equation}
\begin{displaymath}
\Bigl(\gamma_\sigma F_1-\frac{1}{2m_2}\sigma_{\sigma\beta}k_\beta F_2\Bigr)v(q_2)\Bigr]
D_{\mu\sigma}(k)D_{\nu\lambda}(k),
\end{displaymath}
where $p_{1,2}$ and $q_{1,2}$ are four-momenta of the muon and
proton (nucleus) in initial and final states: $p_{1,2}\approx q_{1,2}$. $k$ stands for the 
four-momentum of the exchange photon. The vertex operator describing the photon-proton interaction
is determined by two electromagnetic form factors $F_{1,2}$.
The propagator of exchange photon is taken in the Coulomb gauge.
The lepton tensor $L_{\mu\nu}$ has a completely definite form for each amplitude in Fig.~\ref{fig1}. 
Using the FeynCalc package \cite{fc} we construct the exact expressions for leptonic 
tensors corresponding to muon self-energy and vertex corrections and the correction of the 
amplitude with spanning photon which have the following form:
\begin{equation}
\label{eq:se}
L_{\mu\nu}^{\Sigma}=-\frac{3\alpha}{4\pi}\gamma_\mu(\hat p_1-\hat k)\gamma_\nu\int_0^1\frac{(1-x)dx}{(1-x)m_1^2+x{\bf k}^2},
\end{equation}
\begin{equation}
\label{eq:vertex}
L_{\mu\nu}^{\Lambda}=\frac{\alpha}{4\pi}\int_0^1 dz\int_0^1 dx \gamma_\mu
\frac{\hat p_1-\hat k+m_1}{(p_1-k)^2-m_1^2+i0}
\left[F_\nu^{(1)}+\frac{F_\nu^{(2)}}{\Delta}+\frac{F_\nu^{(3)}}{\Delta^2}\right],
\end{equation}
\begin{equation}
\label{eq:jellyfish}
L_{\mu\nu}^{\Xi}=-2\frac{\alpha}{4\pi}\int_0^1 (1-z)dz\int_0^1 (1-x)x^2dx \left(\frac{F^{(1)}_{\mu\nu}}{\Delta}+
\frac{F^{(2)}_{\mu\nu}}{\Delta^2}+\frac{F^{(3)}_{\mu\nu}}{\Delta^3}\right),
\end{equation}
where the functions $F_\nu^{(i)}$, $F_{\mu\nu}^{(i)}$ are written in explicit form in our previous papers
\cite{apmplb,apm2014}.

To extract the Lamb shift part of the potential, we use special projection operators 
in \eqref{eq:2gamma1}-\eqref{eq:2gamma2} on the states of particles with spin 0 and 1:
\begin{equation}
\label{eq:project}
\hat\Pi_{S=0,1}=[u(p_1)\bar v(p_2)]_{S=0,1}=\frac{1+\hat Q}{2\sqrt{2}}\gamma_5(\hat\varepsilon),~~~Q=(1,0,0,0).
\end{equation}
Inserting \eqref{eq:project} into \eqref{eq:2gamma1}-\eqref{eq:2gamma2} and calculating the trace and contractions over the Lorentz indices by means of the system Form \cite{form} we obtain the contributions to the potential
for states with spin 0 $V(^1S_0)$ and spin 1 $V(^3S_1)$. 
We give here general expressions for the numerators of direct exchange amplitudes for states $^1S_0$ and $^3S_1$:
\begin{equation}
\label{eq:num1}
{\cal N}(^1S_0)=\frac{1}{8}Tr\Bigl\{\gamma_5(1+\hat Q)L_{\mu\nu}(1+\hat Q)\gamma_5
\Bigl[\gamma_\lambda F_1-\frac{1}{2m_2}\sigma_{\lambda\rho}k_\rho F_2\Bigr]
(-\hat p_2-\hat k+m_2)\times
\end{equation}
\begin{displaymath}
\Bigl[\gamma_\sigma F_1+\frac{1}{2m_2}\sigma_{\sigma\beta}k_\beta F_2\Bigr]\Bigr\}
D_{\mu\sigma}(k)D_{\nu\lambda}(k),
\end{displaymath}
\begin{equation}
\label{eq:num2}
{\cal N}(^3S_1)=\frac{1}{24}Tr\Bigl\{\hat\varepsilon^\ast(1+\hat Q)L_{\mu\nu}(1+\hat Q)\hat\varepsilon
\Bigl[\gamma_\lambda F_1-\frac{1}{2m_2}\sigma_{\lambda\rho}k_\rho F_2\Bigr]
(-\hat p_2-\hat k+m_2)\times
\end{equation}
\begin{displaymath}
\Bigl[\gamma_\sigma F_1+\frac{1}{2m_2}\sigma_{\sigma\beta}k_\beta F_2\Bigr]\Bigr\}
D_{\mu\sigma}(k)D_{\nu\lambda}(k),
\end{displaymath}

After that we can present the Lamb shift part
of the potential by means of the following relation \cite{apm2000}:
\begin{equation}
\label{eq:ls_part}
V_{Ls}=V(^1S_0)+\frac{3}{4}V^{hfs}.
\end{equation}
Neglecting the recoil effects we simplify the denominators of the proton propagator as follows:
$1/[(p_2+k)^2-m_2^2+i0]\approx 1/(k^2+2kp_2+i0)\approx 1/(2k_0m_2+i0)$. The crossed two-photon amplitudes 
give a similar contribution to the Lamb shift which is determined also by relations \eqref{eq:2gamma2}-\eqref{eq:jellyfish} with the replacement $k\to -k$ in the proton propagator. 
As a result the summary contribution of direct and crossed amplitudes is proportional to the
$\delta(k_0)$:
\begin{equation}
\label{eq:delta}
\frac{1}{2m_2k_0+i0}+\frac{1}{-2m_2k_0+i0}=-\frac{i\pi}{m_2}\delta(k_0).
\end{equation}
In the case of muonic deuterium \cite{apmplb} the tansformation of the scattering amplitude and a construction
of muon-deuteron potential can be done in much the same way. The main difference is related with
the structure of deuteron-photon vertex functions and projection operators on the states with spin 3/2 and 1/2.
As a result three types of corrections of order  $\alpha(Z\alpha)^5$ to the Lamb shift
in all cases of muonic hydrogen are presented in the integral form over loop momentum ${\bf k}$
and the Feynman parameters. Below, we present the complete integral expressions for the corrections under 
study, as well as the results of analytic integration in the case of the expansion of the form 
factor in a series with preservation of the leading term proportional to $r_N^2$:
\begin{equation}
\label{eq:result1}
\Delta E^{Ls}_{se}=-\frac{6\alpha(Z\alpha)^5\mu^3}{\pi^2m_1^2n^3}\delta_{l0}\int_0^\infty\frac{[G_E^2(k^2)-1]}{k^2(k^2-1)^2}
(1-k^2+2k^2\ln k)dk,
\end{equation}
\begin{equation}
\label{eq:result11}
\Delta E^{Ls}_{se}(r_N^2)=\frac{\alpha(Z\alpha)^5\mu^3r_N^2}{n^3}\frac{1}{2}\delta_{l0},
\end{equation}
\begin{equation}
\label{eq:result21}
\Delta E^{Ls}_{vertex-1}=\frac{48\alpha(Z\alpha)^5\mu^3}{\pi^2m_1^2n^3}\delta_{l0}
\int_0^1 dz\int_0^1 xdx\int_0^\infty
\frac{[G_E(k^2)^2-1]}{k^4}\ln[\frac{x+k^2z(1-xz)}{x}]dk,
\end{equation}
\begin{equation}
\label{eq:result211}
\Delta E^{Ls}_{vertex-1}(r_N^2)=\frac{\alpha(Z\alpha)^5\mu^3r_N^2}{n^3}(1-4\ln 2)\delta_{l0},
\end{equation}
\begin{equation}
\label{eq:result22}
\Delta E^{Ls}_{vertex-2}=\frac{8\alpha(Z\alpha)^5\mu^3}{\pi^2m_1^2n^3}\delta_{l0}\int_0^1 dz\int_0^1 dx
\int_0^\infty\frac{[G_E^2(k^2)-1]}{k^4[x+z(1-xz)k^2]^2}\Bigl[-8x^2+k^2 (-x^2 + 8 z -
\end{equation}
\begin{displaymath}
8 x z - 4 x^2 z + 2 x^3 z - 8 x z^2 + 16 x^2 z^2 - 4 x^3 z^2) + 
k^4 (2 z - 3 x z + 2 x z^2 + 5 x^2 z^2 - 8 x^2 z^3 - 2 x^3 z^3 + 4 x^3 z^4)
\Bigr],
\end{displaymath}
\begin{equation}
\label{eq:result221}
\Delta E^{Ls}_{vertex-2}(r_N^2)=\frac{\alpha(Z\alpha)^5\mu^3r_N^2}{n^3}\Bigl(\frac{4}{3}\ln 2-3\Bigr)\delta_{l0},
\end{equation}
\begin{equation}
\label{eq:result3}
\Delta E^{Ls}_{jellyfish}=\frac{2\alpha(Z\alpha)^5\mu^3}{\pi^2m_1^2n^3}\delta_{l0}\int_0^1(1-z)dz\int_0^1(1-x)dx
\int_0^\infty\frac{[G_E^2(k^2)-1]}{[x+z(1-xz)k^2]^3}dk\times
\end{equation}
\begin{displaymath}
\Bigl[-24x-24x^2+36x^4+k^2(24z-96xz-8xz^2-22x^2z+40x^2z^2+76x^3z+24x^3z^2-48x^4z^2)+
\end{displaymath}
\begin{displaymath}
k^4(-66xz^2+32x^2z^2+70x^2z^3-36x^3z^3-24x^3z^4+12x^4z^4)\Bigr],
\end{displaymath}
\begin{equation}
\label{eq:result31}
\Delta E^{Ls}_{jellyfish}(r_N^2)=\frac{\alpha(Z\alpha)^5\mu^3r_N^2}{n^3}\left(\frac{16}{3}\ln 2-\frac{7}{3}\right)\delta_{l0},
\end{equation}
where an expansion looks as follows:
\begin{equation}
\label{eq:f12}
G_E^2(k^2)-1\approx G_E^2(0)+2G_E(0)G_E'(0)k^2-1=-\frac{1}{3}r_N^2k^2.
\end{equation}
Such an expansion means that one of the vertex operators in Fig.~\ref{fig1} 
is the vertex operator of a point particle, and the other is proportional to $r_N^2$.
The coefficient 2 acts as a combinatorial factor. We note that the vertex correction $\Delta E^{Ls}_{vertex-2}$
includes the contributions of two terms with functions $F_\nu^{(2)}$ and $F_\nu^{(3)}$.
The index "jellyfish" denotes the contribution of the amplitude with spanning photon.

There is another correction of the same order $\alpha(Z\alpha)^5$, determined by the muon vacuum polarization 
effect \cite{egs}. It can be obtained by the same scheme as the previous corrections in Fig.~\ref{fig1}.
To construct the potential, we need to use the following substitution
\begin{equation}
\label{eq:vp}
\frac{1}{k^2}\to \frac{\alpha}{3\pi}\int_1^\infty\frac{\rho(\xi)d\xi}{k^2+4m_l^2\xi^2}
\end{equation}
in one of the exchange photons, $\rho(\xi)=\sqrt{\xi^2-1}(2\xi^2+1)/\xi^4$, $m_l$ is the lepton mass.
After integration over the spectral parameter $\xi$, this correction can be represented 
as a one-dimensional integral ($a_1=2m_l/m_1 $):
\begin{equation}
\label{eq:vp1}
\Delta E^{Ls}_{vp}=-\frac{32\alpha(Z\alpha)^5\mu^3}{3\pi^2m_1^2n^3}\delta_{l0}\int_0^\infty
\frac{[G_E^2(k^2)-1]}{3k^7}\Bigl[3a_1^2k-5k^3-3(a_1^2-2k^2)\sqrt{k^2+a_1^2}\arcsin\frac{k}{a_1}\Bigr].
\end{equation}
After the replacement \eqref{eq:f12} and analytic integration in \eqref{eq:vp1}, we obtain the following result
(the argument given in parentheses denotes the extraction of a contribution proportional to $r_N^2$):
\begin{equation}
\label{eq:vp2}
\Delta E^{Ls}_{vp}(r_N^2)=\frac{\alpha(Z\alpha)^5\mu^3r_N^2}{n^3}\frac{m_1}{2m_l}\delta_{l0}.
\end{equation}
It is clear that we do not consider here the electronic vacuum polarization ($m_l\to m_e$), 
which must be taken into account separately because it presents the correction of a different order.

All corrections \eqref{eq:result1}-\eqref{eq:vp2} are expressed through the convergent integrals.
In the case of expansion \eqref{eq:f12} all integrations can be done analytically.
Some of the integrals contain terms which are divergent at $k=0$ but their sum is finite.
In Table~\ref{t1} we present separate results for muon self-energy, vertex and
spanning photon contributions in the Fried-Yennie gauge. 
We mention here once again that the result of an analytical calculation of the corrections 
from the amplitudes in the case of a point nucleus 
$\Delta E^{Ls}=4(1+11/128-\ln 2/2)\alpha(Z\alpha)^5\mu^3\delta_{l0}/m_1^2n^3 $
was obtained in \cite{kks}.
In paper \cite{eides2} the expressions for the lepton tensors
of the vertex and spanning photon diagrams were constructed in a slightly different form
but they lead to the same contributions to the Lamb shift of $S$-states.
In numerical calculations of integrals \eqref{eq:result1}, \eqref{eq:result21}, \eqref{eq:result22},
\eqref{eq:result3}, \eqref{eq:vp1} with finite-size nucleus we use the known parameterizations 
for electromagnetic form factors of nuclei as in our previous works 
\cite{apmplb,faustov1,faustov2,apm2011}, namely, the dipole parametrization with charge radii from works
\cite{crema1,crema2,crema3,marinova}: $r_p=0.84184\pm 0.00067$ fm, $r_d=2.12562\pm 0.00078$ fm, $r_t=1.7591\pm 0.0363$ fm,  $r_{hel}=1.9661\pm 0.0030$ fm, $r_\alpha=1.6755\pm 0.0028$ fm.
The obtained numerical values are written to four digits after the decimal 
point for the central values of the charge radii. If the accuracy of the results for a deuteron 
and a proton can be considered high (an error of not more than one percent),
then for other nuclei errors in the parametrization of form factors  can be estimated at 5 percent.

\begin{table}[h]
\caption{\label{t1} Radiative nonrecoil nucleus finite size corrections of order $\alpha(Z\alpha)^5$,
to the Lamb shift of $S$-states in light muonic atoms. Numerical results for the ground
state (n=1) are presented. The contribution to the Lamb shift proportional to the nucleus
charge radius squared is indicated in round brackets.}
\bigskip
\begin{tabular}{|c|c|c|c|c|c|}   \hline
Muonic atom  &  SE cor-  &   Vertex cor-   & Spanning photon & VP cor-   &  Summary \\
             & rection,  meV & rection, meV & correction, meV & rection, meV   &  correction, meV   \\    \hline
Finite size       &               &              &                 &                &                    \\
correction  $\sim$  &$\frac{1}{2}$ & $-(\frac{8}{3}\ln 2+2)$ &$(\frac{16}{3}\ln 2-\frac{7}{3})$& $\frac{1}{2}$ &
$(\frac{8}{3}\ln 2-\frac{23}{6}+\frac{1}{2})$    \\    
$ \alpha(Z\alpha)^5\mu^3 r_N^2/n^3$ &    &    &    &    &   \\    \hline
Muonic   &  0.0007  &   -0.0066   &  0.0018  &  0.0005  &   -0.0036  \\
hydrogen     & (0.0012)   &  (-0.0091)    & (0.0032)   & (0.0012)&   (-0.0035)   \\   \hline
Muonic   &  0.0037  & -0.0321    & 0.0087   & 0.0023 & -0.0174   \\
deuterium  & (0.0088)   &(-0.0675)      & (0.0239)  & (0.0088)&   (-0.0260)    \\   \hline
Muonic   &  0.0030  & -0.0274     & 0.0070   & 0.0019&    -0.0155   \\
tritium  & (0.0063)   &(-0.0488)      & (0.0173)  &(0.0063) &   (-0.0188)    \\   \hline
Muonic   &  0.1123  & -1.0548     & 0.2618   &0.0698 &   -0.6109   \\
helium-3  & (0.2532)   &(-1.9492)      & (0.6906)  & (0.2532)&   (-0.7521)    \\   \hline
Muonic   &  0.0899  & -0.8321     & 0.2126   &0.0579 &   -0.4717   \\
helium-4  & (0.1889)   &(-1.4542)      & (0.5152)  &(0.1889) &   (-0.5612)    \\   \hline
\end{tabular}
\end{table}

In the approximation, when the corrections to the structure of the nucleus are proportional 
to the square of the charge radius, our complete analytical result from the Table~\ref{t1} 
coincides with the previous calculations \cite{milstein1,milstein2,egs} (exact analytical
result is written in the book \cite{egs}). Numerical results are obtained with nucleus charge
radii taken from \cite{crema1,crema2,crema3,marinova}.
It follows from obtained results in Table~\ref{t1} that the account of the nucleus structure
by means of electric form factor changes essentially the results in which the expansion
\eqref{eq:f12} is used. For separate contributions, the magnitude of the correction 
decreases more than two times. Such a significant change in the magnitude of the 
corrections is due to an increase in the lepton mass.
If in the case of electron hydrogen the expansion \eqref{eq:f12} works well, then for light 
muonic atoms it is necessary to use the exact integral relations \eqref{eq:result1}, 
\eqref{eq:result21}, \eqref{eq:result22}, \eqref{eq:result3}, \eqref{eq:vp1} obtained 
by us and to carry out calculations using the explicit form of nuclear form factors.
For comparison, we present in the Table~\ref{t1} two numerical results obtained 
on the basis of the calculation of integrals \eqref{eq:result1}, 
\eqref{eq:result21}, \eqref{eq:result22}, \eqref{eq:result3} and analytical formulas 
(in parentheses).
Although in general the corrections obtained are small, they are nevertheless 
considered important when the accuracy of the experiment is increased.
To construct the quasipotential corresponding to
amplitudes in Fig.~\ref{fig1} we develop the method of projection operators on the bound states with
definite spins. It allows to employ different systems of analytical calculations
\cite{form,fc}. In this approach more complicated nuclear structure corrections, 
for example, radiative recoil corrections
to the Lamb shift of order $\alpha(Z\alpha)^5m_1/m_2$ can be evaluated if an increase of the accuracy
will be needed. The results from Table~\ref{t1} should be taken into account to obtain total value 
of the Lamb shift in light muonic atoms for a comparison with experimental data 
\cite{crema1,crema2,crema3,crema4,crema2017,crema2016}.

The authors are grateful to R.~Pohl, F.~Kottmann, J.~J.~Krauth and B.~Franke 
for useful discussions and information about the CREMA collaboration results.
The work is supported by the Russian Foundation for Basic Research (RFBR grant No. 16-02-00554).

\end{document}